P. Belli[1,2], R. Bernabei[1,2], R. S. Boiko[3,4], F. A. Danevich[3], A. Di Marco[1,2], A. Incicchitti[5,6],
D. V. Kasperovych[3], F. Cappella[5], V. Caracciolo[7], V. V. Kobychev[3], O. G. Polischuk[3], N. V. Sokur[3,*],
V. I. Tretyak[3], R. Cerulli[1,2]

*1 INFN, Sezione di Roma "Tor Vergata", Rome, Italy*
*2 Dipartimento di Fisica, Universita di Roma "Tor Vergata", Rome, Italy*
*3 Institute for Nuclear Research, National Academy of Sciences of Ukraine, Kyiv, Ukraine*
*4 National University of Life and Environmental Sciences of Ukraine, Kyiv, Ukraine*
*5 INFN, Sezione di Roma, Rome, Italy*
*6 Dipartimento di Fisica, Universita di Roma "La Sapienza", Rome, Italy*
*7 INFN, Laboratori Nazionali del Gran Sasso, Assergi, Italy*

*Corresponding author: nazar147@ukr.net

# HALF-LIFE MEASUREMENTS OF $^{212}$Po
# WITH THORIUM-LOADED LIQUID SCINTILLATOR[1]

Precise measurement of half-life of $^{212}$Po (one of the daughter nuclides in radioactive chain of $^{232}$Th) was realized by means of liquid scintillator based on toluene doped by complex of thorium and trioctylphosphine oxide with concentration of Th ≈ 0.1 mass %. Fast photomultiplier tube and high frequency oscilloscope were used to acquire the scintillation signals waveforms. The algorithms were developed to find pairs of $^{212}$Bi β-decays and subsequent $^{212}$Po α-decays, to calculate time differences between the events in the pair, and to build $^{212}$Bi β-decay and $^{212}$Po α-decay energy spectra. Preliminary the $^{212}$Po half-life is $T_{1/2} = (294.8 \pm 1.9)$ ns. The experiment is in progress aiming at reduction of the statistical and systematic uncertainties.

*Keywords:* α-decay, $^{212}$Po, half-life.

# ВИМІРЮВАННЯ ПЕРІОДУ НАПІВРОЗПАДУ ЯДРА $^{212}$Po
# ЗА ДОПОМОГОЮ НАСИЧЕНОГО ТОРІЄМ РІДКОГО СЦИНТИЛЯТОРА

Точне вимірювання періоду напіврозпаду ядра $^{212}$Po (один із дочірніх нуклідів радіоактивного ряду $^{232}$Th) було виконано за допомогою рідкого сцинтилятора на основі толуолу, у який було введено торій у концентрації ≈ 0,1 мас. % у вигляді комплексу з оксидом триоктилфосфіну. Для реєстрації сцинтиляційних сигналів було застосовано швидкий фотоелектронний помножувач та високочастотний осцилоскоп. Розроблено алгоритм пошуку подій β-розпадів $^{212}$Bi та наступних за ними α-розпадів $^{212}$Po, визначення часового проміжку між ними, а також побудови спектрів β-розпаду $^{212}$Bi та α-розпаду $^{212}$Po. Попередньо період напіврозпаду ядра $^{212}$Po становить $T_{1/2} = (294,8 \pm 1,9)$ нс. Експеримент триває з метою набрати більшу кількість подій, щоб зменшити статистичну та систематичну похибки.

*Ключові слова:* α-розпад, $^{212}$Po, період напіврозпаду.

---

## 1. Вступ

Радіонуклід $^{212}$Po (дочірній $^{232}$Th) розпадається найшвидше серед природних радіоактивних ядер. Період напіврозпаду $^{212}$Po відносно α-розпаду, що з вірогідністю 100 % іде на основний стан $^{208}$Pb, становить усього $T_{1/2} = 299 \pm 2$ нс [1]. Нещодавно було виконано кілька експериментів, в яких було виміряно період напіврозпаду ядра $^{212}$Po: за допомогою рідкого сцинтилятора детектора Borexino $T_{1/2} = [294{,}7 \pm 0{,}6$ (стат) $\pm 0{,}8$ (сист)] нс [2], із сцинтилятором фториду барію $T_{1/2} = [298{,}8 \pm 0{,}8$ (стат) $\pm 1{,}4$ (сист)] нс [3] та в експерименті з пошуку темної матерії XENON $T_{1/2} = [293{,}9 \pm 1{,}0$ (стат) $\pm 0{,}6$ (сист)] нс [4]. Недоліком експериментів [2] і [4] є застосування детекторів великого об'єму і великої кількості фотоелектронних помножувачів (ФЕП) з невисокими часовими характеристиками, у той час як сцинтилятор фториду барію, використаний у вимірюваннях [3], має значно повільніший відгук у порівнянні з рідкими сцинтиляторами. У даному експерименті було поставлено задачу покращити точність визначення періоду напіврозпаду ядра $^{212}$Po з використанням рідкого сцинтилятора малого об'єму (для зменшення розкиду часу надходження сцинтиляційних сигналів), який би проглядався одним ФЕП з якомога вищими часовими характеристиками.

## 2. Розробка насиченого торієм рідкого сцинтилятора

Щоб виготовити насичений торієм рідкий сцинтилятор, було використано найбільш поширену сполуку з торієм – торію нітрат пентагідрат Th(NO$_3$)$_4 \cdot$ 5H$_2$O. Зважаючи на те, що нітрат торію не розчиняється в неполяризованих органічних розчинниках, які використовують для рідких сцинтиляторів, було застосовано комплекс торію з триоктилфосфін оксидом (TOPO). Було отримано 20 % розчин TOPO в толуолі, який був змішаний з нітрат пентагідратом торію. Цей розчин містив 2 мг торію в 1 мл TOPO (приблизно 5 мг Th(NO$_3$)$_4 \times$ $\times$ 5H$_2$O в 1 мл). Розчинення нітрату торію можна схематично зобразити як

$$\text{Th}(NO_3)_4 \cdot 5H_2O + 3TOPO_{(liquid)} \rightarrow$$

$$\rightarrow \text{Th} \cdot 3TOPO(NO_3)_{4(liquid)} + 5H_2O.$$

Отриманий органічний розчин із торієм було розчинено в рідкому сцинтиляторі, що являв собою розчин 0,1 % 2,5-дифенілоксазолу (C$_{15}$H$_{11}$NO, сцинтилююча добавка) та 0,01 % 5-феніл-2-[4-(5-феніл-1,3-оксазол-2-іл)феніл]-1,3-оксазолу (C$_{24}$H$_{16}$N$_2$O$_2$, зміщувач спектра сцинтиляційного випромінювання) у толуолі. Таким чином, концентрація торію в рідкому сцинтиляторі (PC:Th) мала становити близько 0,1 мас. %.

Збагачений торієм рідкий сцинтилятор має містити $^{232}$Th, $^{228}$Th та їхні дочірні елементи, у тому числі $^{212}$Bi і $^{212}$Po. Ми припускаємо, що активність $^{228}$Ra в сцинтиляторі має бути достатньо низькою у зв'язку з хімічними процедурами підготовки (Th · 3TOPO)(NO$_3$)$_4$ (які мали вилучити радій) та порівняно великим періодом накопичення $^{228}$Ra (кілька років) завдяки розпаду $^{232}$Th. Також ми припускаємо низьку активність у сцинтиляторі урану і його дочірніх нуклідів, зокрема $^{226}$Ra (події швидкого ланцюжка розпадів $^{214}$Bi $\rightarrow$ $^{214}$Po $\rightarrow$ $^{210}$Pb з періодом напіврозпаду $164{,}3 \pm 2{,}0$ мкс можуть заважати точному визначенню періоду напіврозпаду $^{212}$Po).

## 3. Експеримент

### 3.1. Визначення активності торію у рідкому сцинтиляторі

Для визначення активності торію та його дочірніх нуклідів 7 мл PC:Th було поміщено у кварцову кювету і встановлено на ФЕП Philips XP2412. Енергетичний спектр показано на рис. 1.

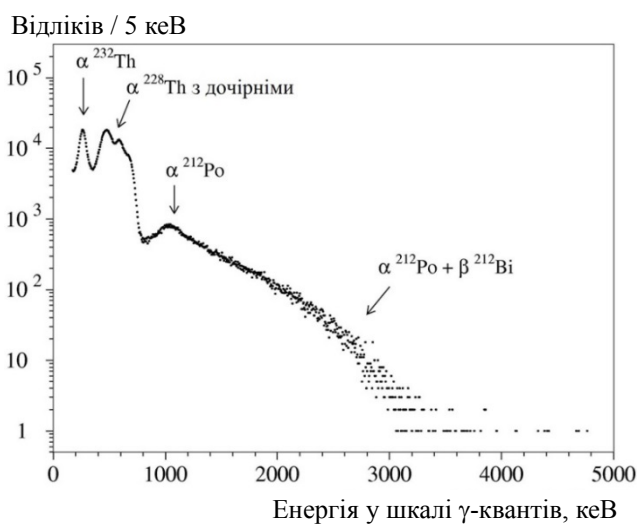

Рис. 1. Енергетичний спектр, накопичений із PC:Th впродовж 5910 с. Показано α-піки $^{232}$Th, $^{228}$Th та їхніх дочірніх нуклідів, а також розподіли від накладання α і β подій від ланцюжка розпадів $^{212}$Bi $\rightarrow$ $^{212}$Po $\rightarrow$ $^{208}$Pb.

Зі спектра (див. рис. 1) було оцінено активність 232Th і 228Th у сцинтиляторі 4,61 і 3,82 Бк/мл (активність $^{228}$Th, оскільки вона змінюється у часі, наведено на 7 липня 2016 р.), що відповідає концентрації торію 0,113(1) мас. %, відповідно до процедури приготування сцинтилятора. Крім того, із цих вимірювань було визначено залежність так званого α/γ співвідношення від енергії α-час-тинок: α/γ = 0,02149(14) + 0,1104(3) · $10^{-4}$ · $E_\alpha$ ($E_\alpha$ – енергія α-частинок, кеВ).

### 3.2. Вимірювання форми сцинтиляційних сигналів

5 мл PC:Th було поміщено у кварцову кювету, установлену на швидкий фотоелектронний помножувач Hamamatsu R13089-100-11 із часом наростання сигналу 0,9 нс та розкидом часу поширення фотоелектронів від катоду до аноду 0,17 нс (ширина піка на половині висоти). Форми сигналів записувалися за допомогою високочастотного осцилоскопа LeCroy WavePro 735Zi-A з частотою запису форм 20 Гз/с (відповідно ширина часового каналу становила 50 пс) і смугою пропускання 3,5 ГГц. Усього було записано 785548 подій, з яких було відібрано 69254 BiPo-подій β-розпадів $^{212}$Bi з наступними за ними α-розпадами $^{212}$Po (8,82 % від усіх подій).

### 4. Аналіз даних

#### 4.1. Алгоритми пошуку BiPo-подій

Типовий сцинтиляційний сигнал PC:Th показано на рис. 2.

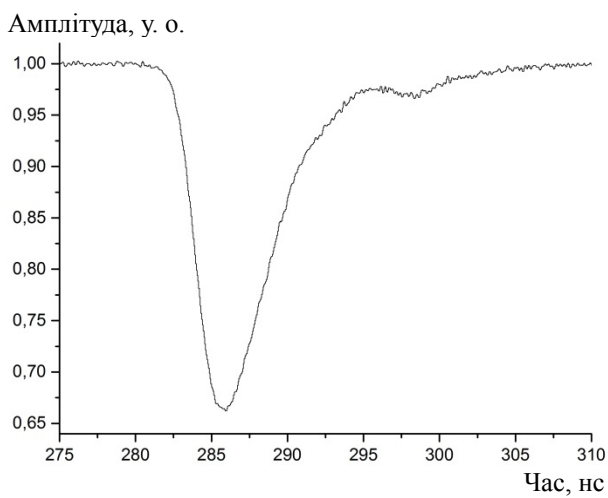

Рис. 2. Приклад форми сцинтиляційного сигналу в рідкому сцинтиляторі. Приблизно через 12 - 14 нс після початку сигналу спостерігається післяімпульс, пов'язаний з пружним розсіянням електронів на першому диноді фотоелектронного помножувача.

Видно, що сигнал приблизно через 12 - 14 нс супроводжується післяімпульсом, який виникає в результаті пружного розсіяння електронів на першому диноді фотоелектронного помножувача. Це значно ускладнює дослідження BiPo-подій у випадках, коли час між сигналами порівняний або менший за час спаду амплітуди післяімпульсу (менше 30 нс). По-перше, ми не можемо якісно оцінити початок другого сигналу, по-друге, наявність післяімпульсу веде до збільшення площі другого сигналу, що спотворює енергетичний спектр α-розпаду $^{212}$Po.

Визначення будь-яких особливостей сигналу з математичної точки зору є простим дослідженням функції. Алгоритм дослідження функції такий. Спочатку ми задали точність розрахунків. Знаючи приблизно час початку першого сигналу (≈ 280 нс), ми починали збирати інформацію про базову лінію від цього часу у від'ємному напрямку, щоб визначити параметри базової лінії недалеко від сигналу. Критерій закінчення збору інформації про базову лінію був такий: ⟨U(n + 1)⟩ - ⟨U(n)⟩ < 0,01 · σ(n), де ⟨U(n + 1)⟩ – середня амплітуда (n + 1) кількості каналів, ⟨U(n)⟩ – середня амплітуда (n) кількості каналів, σ(n) – середнє квадратичне відхилення (n) кількості каналів. Якщо умова не виконувалася, то до вже оброблених каналів додавався наступний й обраховувалися нові параметри базової лінії, тоді умова перевірялася знову. Таким чином, на виході ми отримали характеристики базової лінії: середнє значення та середнє квадратичне відхилення (рис. 3). Знайдене середнє квадратичне відхилення ми використали як точність у дослідженні функції. Кожен сигнал було згруповано у групи з 10 послідовних каналів (часовий інтервал кожної групи 0,5 нс). У кожному циклі дослідження сигналу використовувалися три послідовні групи (приклад таких груп показано на рис. 3). Власне, щоб дослідити сигнал, ми використали так звану Finite-State Machine (FSM). FSM – це абстрактна машина, що перебуває в одному зі своїх заданих станів у будь-який момент часу. Наша FSM мала такі стани: 1) базова лінія; 2) спад сигналу; 3) зростання сигналу; 4) постійна (мається на увазі, що значення функції є або константою, або флуктує навколо певного значення). Певні команди переводять FSM з одного стану в інший, а саме: а) вниз; б) вгору; в) прямолінійне коливання; г) коливання вздовж базової лінії; д) вихід на константу (означає, що сигнал досяг насичення). Перехід зі стану "i" функції в стан "k" називатимемо особливою точкою типу "ik". Особлива точка, де відбувається перехід зі стану 3) у стан 2), – це максимум функції.

Таким чином, проаналізувавши функцію з використанням FSM, ми мали всі необхідні дані, щоб відшукати потенційні BiPo-події. За необхідний нам корисний сигнал ми вважали такий, що задовольняє дві умови.

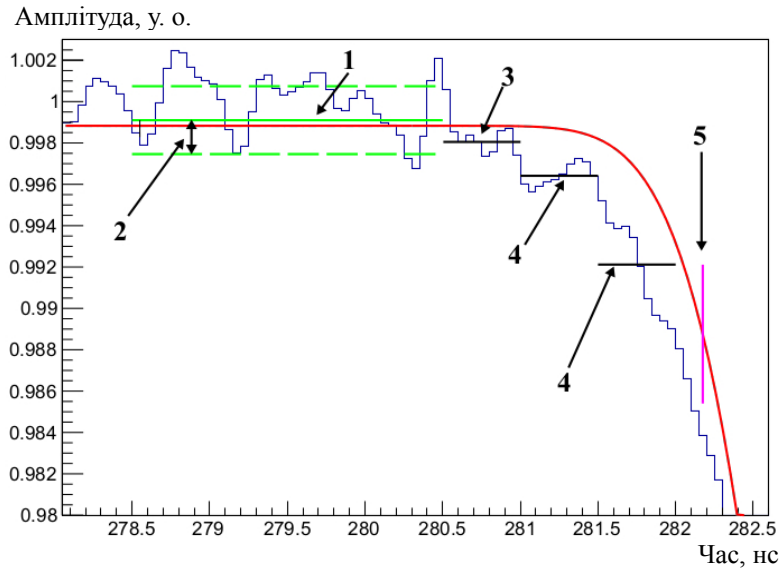

Рис. 3. Апроксимація сцинтиляційного сигналу однією функцією імпульсу (суцільна крива). *1* – середній рівень базової лінії (суцільна пряма лінія); *2* – середнє квадратичне відхилення базової лінії; *3* – група з 10 каналів, в якій середнє значення амплітуд відрізняється від базової лінії більш ніж на половину середнього квадратичного відхилення базової лінії; *4* – групи каналів, в яких середнє значення амплітуд перевищує середнє значення попередньої групи на величину, більшу за одне середнє квадратичне відхилення базової лінії (ознака початку сигналу); *5* – положення в часі на функції імпульсу (прийняте за початок сигналу), де її величина становить 3 % від її максимального значення, яке досягається при ≈286 нс (див. рис. 2). Видно, що апроксимаційна функція зміщена відносно сигналу.

1. Відстань від особливої точки на базовій лінії (в якій сигнал починає зростати, початок сигналу) до наступної особливої точки після максимуму сигналу (коли сигнал знову починає наростати за рахунок післяімпульсу) перевищує 8 нс. Цей часовий інтервал ми вважали за довжину сигналу. Такий часовий інтервал обрано з міркувань, щоб післяімпульс не впливав на ідентифікацію корисного сигналу (обмеження зверху) і щоб відкидати шумові сигнали фотоелектронного помножувача, що характеризуються суттєво меншою довжиною близько 3 - 4 нс (обмеження знизу).

2. Різниця значень функції в точці початку сигналу і в точці максимуму сигналу (цю різницю ми вважали за амплітуду сигналу) перевищує 20 середньоквадратичних відхилень базової лінії. Таким чином, відкидалися сигнали малої амплітуди, післяімпульси, шумові імпульси різної природи тощо.

У результаті такого аналізу ми отримували всю необхідну інформацію про сигнал. На основі цієї інформації задавалися початкові параметри апроксимаційної функції. Апроксимація сигналу виконувалася в межах $[t_0 - 5; t(1/3 \cdot U)]$ нс, де $t_0$ – часовий канал, який було визначено як початок сигналу; $t(1/3 \cdot U)$ – канал, де амплітуда сигналу спала до 1/3 від свого максимального значення. Такий вибір меж апроксимації було обумовлено необхідністю надійного визначення базової лінії, а також уникнення впливу післяімпульсів (включення яких у процедуру підгонки вимагає суттєвого ускладнення апроксимуючої функції).

Спочатку ми спробували апроксимувати сигнали однією функцією (назвемо її функцією імпульсу):

$$f(t) = y_0 + \frac{A}{\sqrt{\frac{\tau_1}{p}\tau_2}}\left(1 - e^{-\frac{t-t_0}{\tau_1}}\right)^p \cdot e^{-\frac{t-t_0}{\tau_2}}, \quad (1)$$

де $y_0$ – постійна, яка описує базову лінію; $A$ – величина, пропорційна амплітуді сигналу; $\tau_1$ – постійна наростання сигналу; $\tau_2$ – постійна спаду сигналу; $p$ – параметр для згладжування початку сигналу. Вибір функції обумовлений тим, що кінетика сцинтиляційних сигналів та перехідні процеси в аналогових радіоелектронних пристроях (зокрема, підсилювачах сигналів) описуються експоненційними функціями. При цьому за початок сигналу ми брали час, де амплітуда апроксимаційної функції досягає 3 % від її максимального значення (по модулю, див. рис. 3). Варто зауважити, що рівень 3 % (трохи вище рівня шуму) було вибрано, щоб відкинути шумові імпульси і водночас відібрати сигнали з достатньо низьким порогом. Останнє було важливо для побудови й дослідження спектра β-частинок $^{212}$Bi. Розробка опти-мального методу визначення початку сигналів триває й очікується, що такий метод буде застосовано для остаточного аналізу даних експерименту. За результатами апроксимації BiPo-сигналів однією функцією було отримано значення часу наростання першого (другого) сцинтиляційного сигналу $\tau_1 = 2{,}0 \pm 0{,}3$ нс ($\tau_1 = 1{,}6 \pm 0{,}2$ нс) і значення постійної спаду першого (другого) сигналу $\tau_2 = 2{,}5 \pm 0{,}3$ нс ($\tau_2 = 3{,}0 \pm 0{,}3$ нс). Але на рис. 3 видно, що така апроксимація неточно описує початкову частину сигналу. По-перше, апроксимаційна функція зміщена відносно сигналу в бік більшого часу. По-друге, постійна, що має описувати базову лінію, також відхиляється від виміряного положення базової лінії. Тому було прийнято рішення застосувати дві функції (1) для опису сигналу за таким алгоритмом:

1) спочатку сигнал апроксимується однією функцією;

2) значення параметрів функції з кроку 1 задаються як початкові параметри однієї з комбінацій двох функцій імпульсу. В іншій (допоміжній) функції імпульсу задавалися певні, однакові для всіх сигналів, початкові значення параметрів;

3) повторювалася апроксимація з комбінацією двох функцій імпульсу (але з однією постійною $y_0$ для опису базової лінії).

Задачею допоміжної функції було компенсувати недоліки на тих ділянках сигналу, де не вистачало основної функції. В обох випадках для мінімізації застосовувався алгоритм Migrad із програмного пакета Minuit [5]. Функцією мінімізації була $\chi^2$-функція, в якій за похибку кожного значення у каналі було прийнято середнє квадратичне відхилення базової лінії. Результат опису сигналу двома функціями показано на рис. 4.

Варто відзначити, що допоміжна функція була необхідною лише для опису початку сигналу. Ми поки що не можемо пояснити причину такої форми сигналів, але вона краще їх описує. Крім того, амплітуда першої функції, як правило, є невеликою, як це можна бачити на рис. 5.

### 4.2. Енергетичні спектри подій розпадів $^{212}$Bi та $^{212}$Po

Коли апроксимацію сигналу зроблено, можна проінтегрувати функцію, яка його описує, у певних межах та знайти амплітуду події у відносних одиницях. Межі інтегрування визначалися в точках, де амплітуда становила 1 % від максимуму функції. При такому виборі меж інтегрування інтегрується більша частина функції, що описує сигнал. Це важливо для зниження впливу статистичних флуктуацій кількості фотоелектронів, а отже, для досягнення якнайвищої енергетичної роздільної здатності сцинтиляційного детектора.

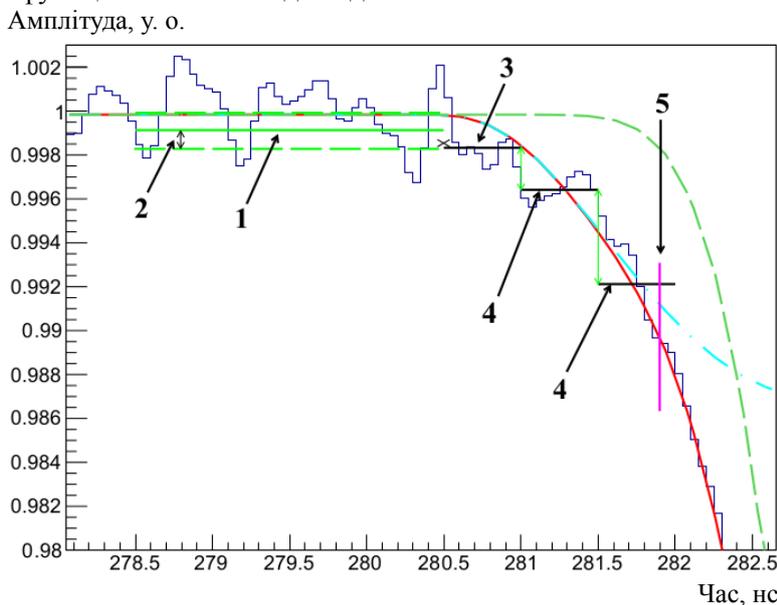

Рис. 4. Опис сигналу за допомогою двох функцій імпульсу. Апроксимацію сигналу показано суцільною лінією, функції – штриховою (основна) і штрих-пунктирною (допоміжна) лініями. *1* – середній рівень базової лінії; *2* – середнє квадратичне відхилення базової лінії; *3* – група з 10 каналів на початку сигналу, в якій середнє значення відрізняється від базової лінії більше, ніж на половину середнього квадратичного відхилення; *4* – групи каналів, в яких середнє значення амплітуд перевищує середнє значення попередньої групи на величину більшу за одне середнє квадратичне відхилення; *5* – час на функції імпульсу, де її величина становить 3 % від її максимального значення, яке досягається при ≈286 нс (див. рис. 2). Цей час прийнято за початок сигналу.

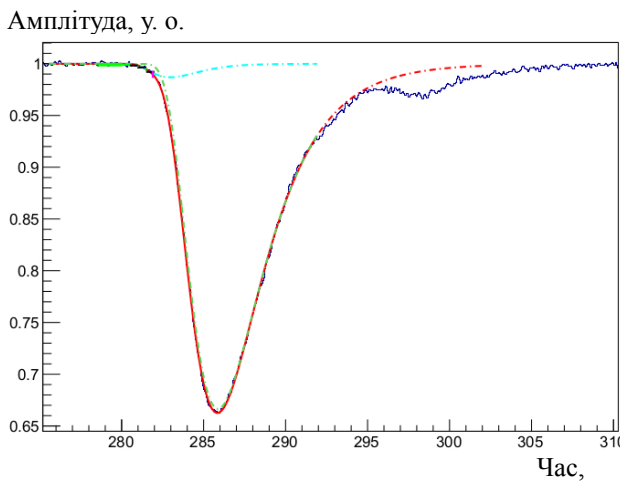

Рис. 5. Сцинтиляційний сигнал, апроксимований двома функціями. Допоміжна функція показана штрих-пунктирною лінією на початку сигналу.

З іншого боку, вибір занадто великого інтервалу інтегрування призведе до збільшення впливу шумів на енергетичну роздільну здатність. Розробка методів аналізу даних триває і оптимальний часовий інтервал для побудови енергетичних спектрів подій β-розпаду $^{212}$Bi та α-розпаду $^{212}$Po буде застосовано при остаточному аналізі даних. β-спектр $^{212}$Bi та α-спектр $^{212}$Po показано на рис. 6.

З виразу для α/γ співвідношення, наведеного в розділі 3.1, можна обрахувати, що максимум α-піка має бути на енергії близько 1040 кеВ в енергетичній шкалі, виміряній з γ-квантами.

Як видно з рис. 6, низькоенергетична частина β-спектра $^{212}$Bi не реєструвалася через встановлений поріг реєстрації осцилоскопа. Події з малою амплітудою (близько 0,15 у. о.) в α-спектрі $^{212}$Po найімовірніше є результатом помилкового визначення післяімпульсів як других сигналів.

### 4.3. Визначення періоду напіврозпаду $^{212}$Po

Спектр часових інтервалів між першим сигналом β-розпаду $^{212}$Bi та другим сигналом α-розпаду $^{212}$Po показано на рис. 7.

Для визначення періоду напіврозпаду $^{212}$Po спектр було апроксимовано експоненційною функцією в різних часових інтервалах (нижня межа змінювалася від 30 до 100 нс, а верхня від 1000 до 1200 нс) та при різному групуванні каналів у спектрі часових інтервалів (1, 2, 4, 5 та 10 нс). Значення $\chi^2$/n.d.f. апроксимацій коливалося в межах 293,25 - 296,00 нс із серенім значенням 294,78 ± 1,59 (стат) нс.

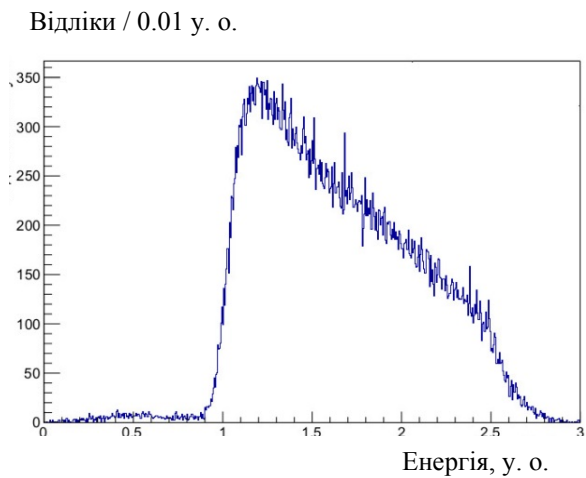 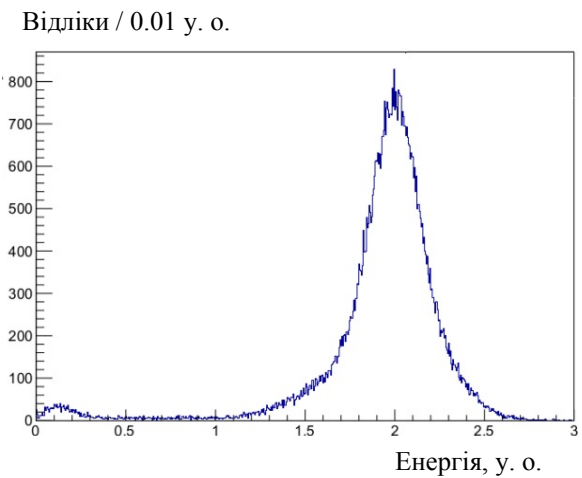

Рис. 6. -спектр 212Bi (ліворуч) та -спектр 212Po (праворуч).
Поріг реєстрації подій осцилоскопом було встановлено на рівні близько 1 у. о.

Внесок у систематичну похибку меж апроксимації дорівнює 0,65 нс; внесок, пов'язаний з групуванням каналів, становить усього 0,002 нс; внесок ширини каналу осцилоскопа 0,05 нс; внесок розкиду часу поширення фотоелектронів від катоду до аноду 0,06 нс; помилка внаслідок неточності визначення початку сигналів оцінена на рівні 0,25 нс. Сумуючи консервативно всі помилки лінійно, отримаємо систематичну похибку 1,0 нс. Таким чином, період напіврозпаду $^{212}$Po становить $T_{1/2} = $ = [294,8 ± 1,6 (стат) ± 1,0 (сист)] нс. Сумуючи статистичну і систематичну помилки квадратично, отримаємо значення $T_{1/2} =$ (294,8 ± 1,9) нс.

### 5. Висновки

Період напіврозпаду ядра $^{212}$Po виміряно за допомогою спеціально розробленого рідкого сцинтилятора на основі толуолу, насиченого торієм з активністю ($^{232}$Th) 4,61 Бк/мл. Сцинтиляційні сигнали реєструвалися за допомогою фотопомножувача із часом наростання сигналу 0,9 нс та розкидом часу поширення фотоелектронів від катоду до аноду 0,13 нс (ширина піка на половині висоти) і

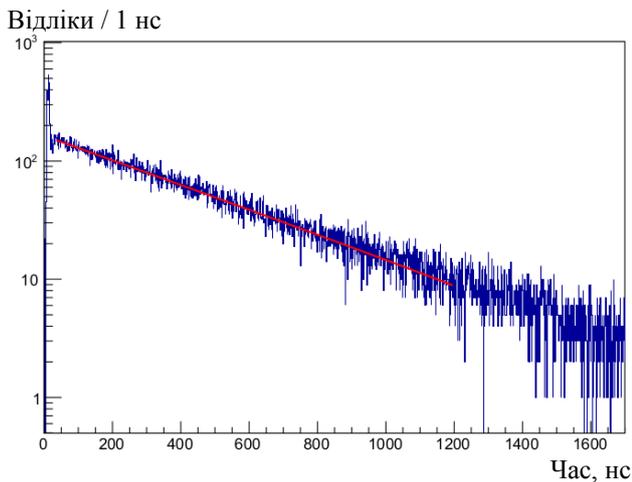

Рис. 7. Спектр часових інтервалів між сигналом β-розпаду ²¹²Bi та сигналом α-розпаду ²¹²Po. Апроксимацію розподілу експоненційною функцією, що відповідає періоду напіврозпаду ядра ²¹²Po $T_{1/2}$ = = 294,8 нс, показано суцільною лінією.

високочастотного осцилоскопа з частотою запису форм 20 Гз/с і смугою пропускання 3,5 ГГц. Виміряний у даному експерименті період напіврозпаду ядра ²¹²Po $T_{1/2}$ = (294,8 ± 1,9) нс є точнішим за табличне значення $T_{1/2}$ = (299 ± 2) нс, узгоджується з останніми вимірюваннями, виконаними колаборацією Borexino – $T_{1/2}$ = [294,7 ± 0,6 (стат) ± ± 0,8 (сист)] нс та колаборацією XENON – $T_{1/2}$ = = [293,9 ± 1,0 (стат) ± 0,6 (сист)] нс, і дещо відрізняється від результату вимірювань із сцинтилятором фториду барію $T_{1/2}$ = [298,8 ± 0,8 (стат) ± ± 1,4 (сист)] нс. Триває аналіз даних нового етапу експерименту з метою отримати більшу кількість подій і таким чином зменшити статистичну та систематичну похибки значення періоду напіврозпаду ядра ²¹²Po.